\documentclass[seceq]{ptptex}

\usepackage{graphicx}
\usepackage{wrapft}

\newcommand{\be}{\begin{eqnarray}}
\newcommand{\ee}{\end{eqnarray}}

\title{Emerging Theory of
Strongly Coupled Quark-Gluon Plasma
}

\author{
Edward \textsc{Shuryak}$^{1,}$\footnote{ e-mail address:
shuryak@tonic.physics.sunysb.edu}  
}

\inst{
$^1$ 
 Department of Physics and Astronomy, University at Stony Brook,\\
Stony Brook NY 11794 USA
}

\abst{
 RHIC data have shown robust collective flows, including
 recent spectacular ``conical flow'' from quenched jets: 
 that confirms that QGP above the
critical line is in a strongly coupled
 regime. One way to study
Non-Abelian classical strongly coupled plasmas is
 via molecular dynamics, which was recently extended
to plasmas with electric and magntic charges.
First results
on its transport (diffusion and viscosity) are reported.
 AdS/CFT correspondence is another actively pursuit
approach to strong coupling regime: we review new results for
  heavy quark
motion and even complete picture of
conical flow, obtained from linearized gravity dual.
 Recent developments in (non-linearized) gravity aims at
 reproduction of the whole time-dependent picture
of the explosion. We finally compare
 transport properties obtained from RHIC data to those
obtained in MD and AdS/CFT.
}

\begin{document}

\maketitle
\section{Why strongly coupled?}

 A realization~\cite{Shu_liquid,SZ_rethinking,SZ_CFT} that
QGP at RHIC  is not a weakly coupled 
gas but rather a strongly coupled liquid
 has lead to a  paradigm
 shift in the field. It was extensively debated  at
the ``discovery''  BNL workshop in 2004
and multiple other meetings since.
The experimental situation was then summarized by ``white papers''
of four RHIC experiments, who basically confirmed
this picture.
In the last 4 years strong efforts has been made to
understand  $why$ QGP (at least) at $T=(1- 2)T_c$
is strongly coupled, and what exactly it means. Another
formidable ``old'' problem is that of confinement/deconfinement,
which got new attention lately.
We have learned a lot
 from other branches of physics  which
had  experience with strongly coupled systems:
atomic gases in strongly coupled
regime, classical plasmas and AdS/CFT.
Those provided important clues: but we are  somewhere in the middle
of the process, just starting to see
how  it all makes a common picture.

%
%
The list of arguments explaining $why$ we think  QGP
is strongly coupled at $T>T_c$
is long and constantly growing:\\
\noindent
1.{\bf Phenomenology}: Collective flows observed at RHIC lead
 hydro practitioners to a conclusion that
 QGP as a ``near perfect liquid'', with unexpectedly
small {\em viscosity-to-entropy ratio}
$\eta/s=.1-.2<<1$ 
in striking contrast to pQCD predictions.
Charmed and possibly even b quarks are strongly quenched:
their diffusion constant $D_c$ (deduced from the data
on single electron quenching and  elliptic flow)  
is much lower than pQCD expectations.
\\
2. {\bf Lattice/spectroscopy}: Lattice data suggest rather heavy quasiparticles and
strong interparticle
potentials, combining the two one
  finds a lot of quasiparticle bound states
\cite{SZ_rethinking}. 
That is why $\eta_c,J/\psi$ remain bound
at $T=(1-2)T_c$, as found 
on the lattice
\cite{charmonium} (and also at RHIC).
Heavy-light resonances help to explain charm stopping
 \cite{SZ_rethinking,Rapp_vanHees}. Near-zero bound states
(Feshbach-type resonances) are known to turn
dilute ultracold trapped atoms into a strongly coupled liquid
with small viscosity \cite{Gelman:2004fj}. 
\\
3.{\bf Classical plasmas}: The interaction parameter   
$\Gamma\sim {<pot.energy>\over <kin.energy>}$,
is not small in sQGP. Classical electromagnetic plasmas  at  comparable
coupling $\Gamma\sim 1-10$ are also good liquids.
This is also true for non-Abelian
plasmas\cite{GSZ}, as well as plasmas containing magnetic monopoles
\cite{Liao:2006ry}.
\\
4. {\bf AdS/CFT} correspondence between conformal field theory
(CFT) $\cal N$=4 supersymmetric YM
at strong coupling and string theory in Anti-de-Sitter
space (AdS) in weak coupling 
  is the basis for many intriguing results
on the CFT plasma properties. Those  are generally 
  close to what is observed 
for sQGP: both 
 are good liquids with record low viscosity, 
which strongly quench heavy quarks,
generate conical flow and have rapid onset of hydro regime.
\\
5.{\bf Electric-Magnetic duality} is perhaps the key to
 confinement. In
the ``Seiberg-Witten'' theory \cal{N}=2 SUSY YM 
confinement is induced by 
 monopole condensation.
  In QGP, as $T$ decreases toward $T_c$,
one finds rapid activation of magnetic monopoles.
 U(1) beta function demands they
 are $weakly$ coupled in IR.
Then, due to  Dirac condition,
 electric coupling is forced to be $strongly$ coupled.
 Uncondensed monopoles\cite{Liao:2006ry} seem to be
an important player in the sQGP close to $T_c$.

\section{Collective Flows in Heavy Ion Collisions}

\begin{figure}
 \includegraphics[width=6cm]{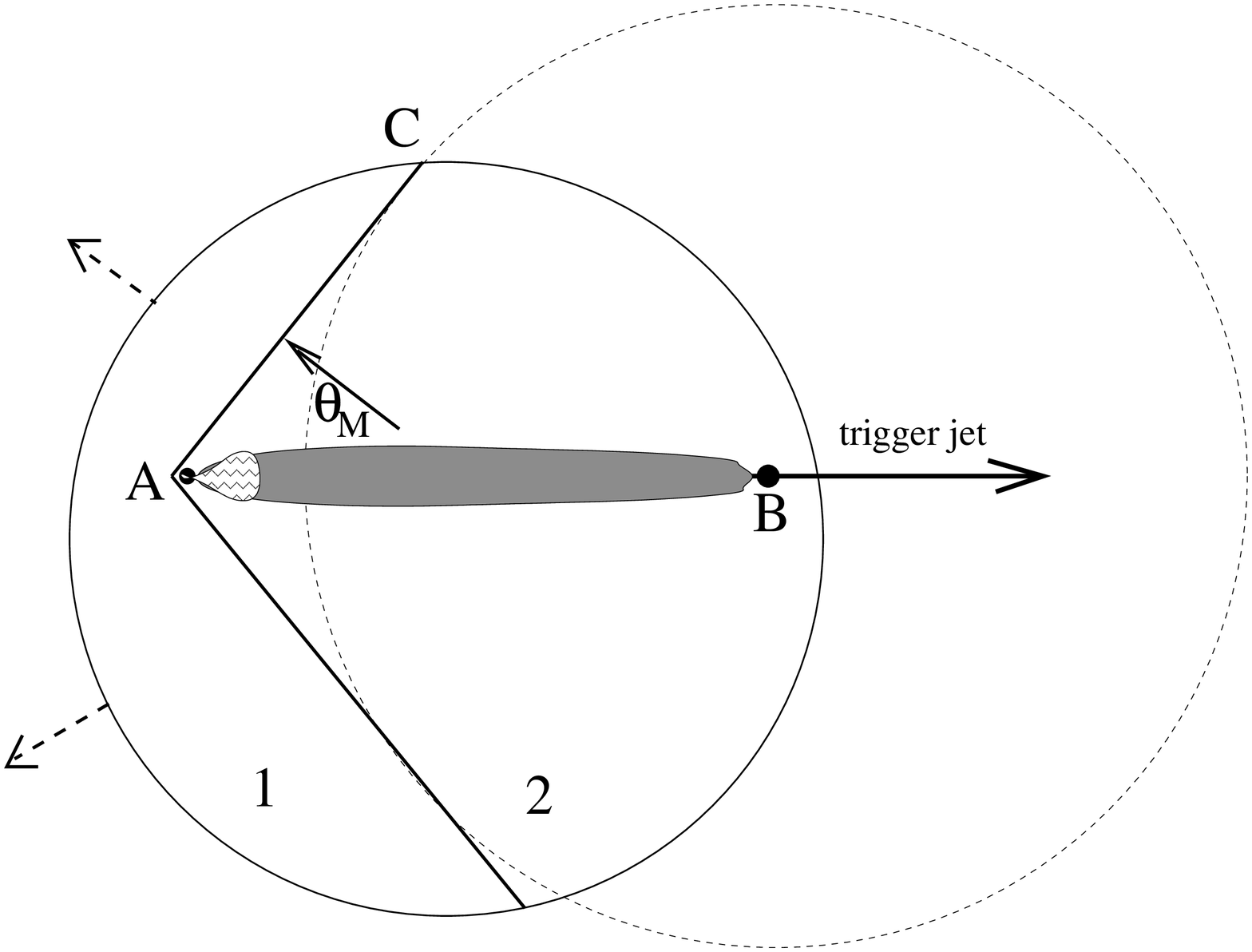}
 \includegraphics[width=5cm]{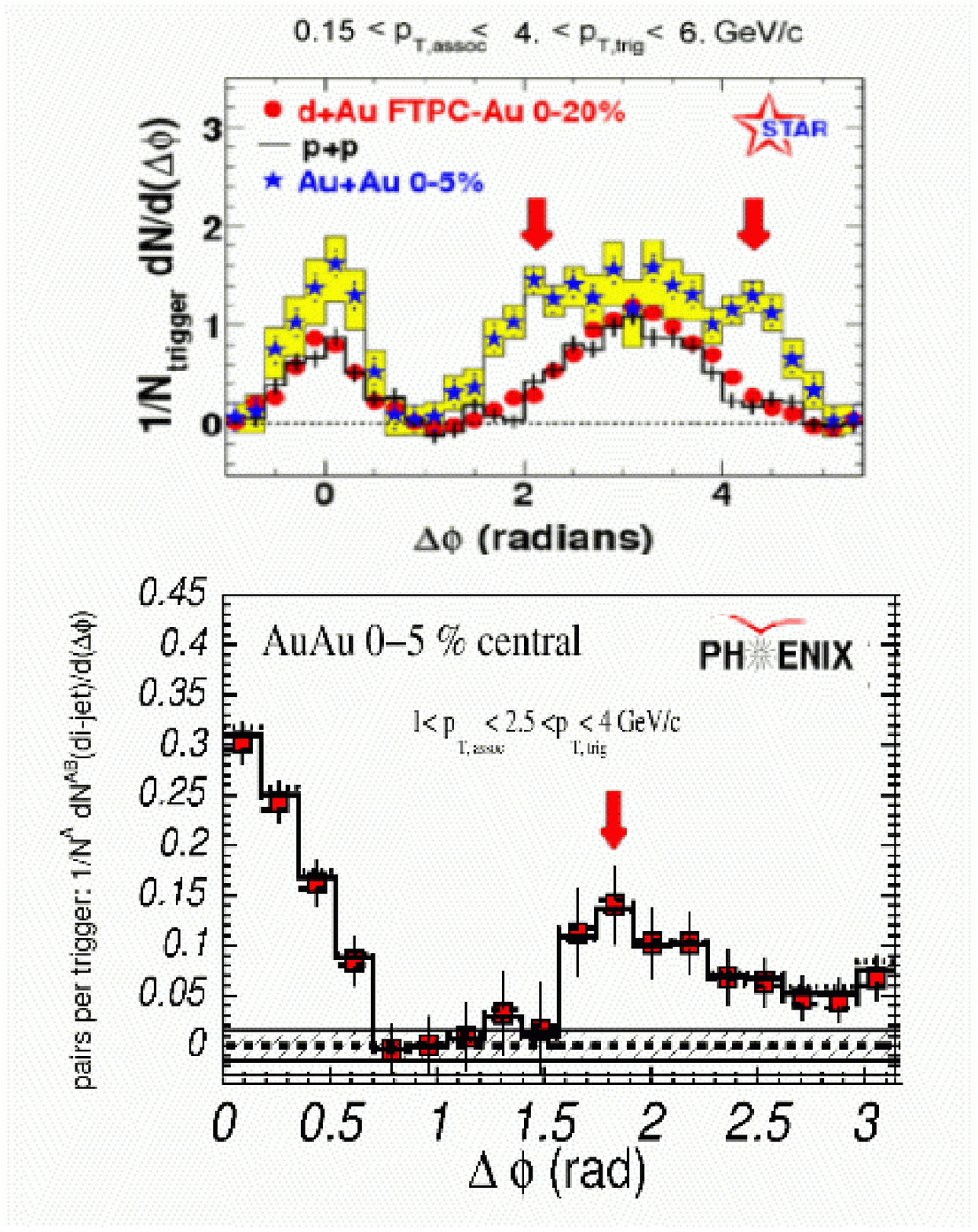}
 \caption{\label{fig_shocks}
(a) A schematic picture of flow created by a jet going through
the fireball. The trigger jet is going to the right
from the origination point  B. The
  companion quenched jet is moving to the left, heating the matter
(in shadowed  area) and producing a shock cone
with a flow normal to it, at
the Mach angle $cos\theta_M=v/c_s$, where $v,c_s$ are jet and sound velocities.
(b)The background subtracted correlation functions from STAR and PHENIX
experiments, a distribution in azimuthal
angle $\Delta\phi$ between the trigger jet and associated particle.
Unlike in pp and $dAu$ collisions where the decay of the companion jet
create a peak at  $\Delta\phi=\pi$ (STAR plot), central $AuAu$
collisions show a minimum at that angle and a maximum corresponding
to the Mach angle (downward arrows).
}
 \end{figure}

Collective radial and elliptic
 flows, related with explosive behavior of hot
matter, 
 observed at SPS and RHIC, are quite accurately reproduced
by the ideal hydrodynamics.
%
  The so called {\em conical} flow\cite{CST} is a 
 hydrodynamical phenomenon 
 induced by
jets quenched in sQGP. 
  Fig.\ref{fig_shocks}(a) explains a view of the process,
in a plane transverse to the beam.  Two oppositely
moving jets originate from   the hard collision point B.
 Due to strong quenching, the survival of the trigger
jet biases it to be produced close to the surface and to
 move outward. This  forces its companion to 
move inward through matter and to be maximally quenched.
The energy deposition starts at point B, thus a spherical sound wave
appears (the dashed circle in Fig.\ref{fig_shocks} ). Further 
 energy deposition is along the jet line, and is propagating with a speed of
light, 
till the leading parton is found at point A
at the moment of the snapshot.
The main prediction is that 
associated secondaries 
  fly preferentially to a very large angle
 $\approx 70$ degrees relative to jet, which is
 consistent with the Mach angle for (
a time-averaged) speed of sound.
As shown in Fig.\ref{fig_shocks}(b), this seem to be what indeed is
observed. Its studies at RHIC 
has been extended to 3-particle correlations, 
which confirmed conical structure of the effect.
The 2-particle signal
for conical flow has been reported at SPS by CERES collaboration
(see proc. of QM06). These observation further proves that
viscosity of the produced matter is small enough, allowing
these waves too survive till freezout time and be observed in spectra.

Antinori and myself\cite{Antinori:2005tu} suggested
that  b-quark jets,
which can be tagged experimentally, will
further test that 
  the angle depend on velocity, not momentum, of the jets:
the cone should then shrink to zero angle at 
$v=c_s=1/\sqrt{3}$.
Casalderrey and myself\cite{CS_variable} have shown, using 
conservation of adiabatic invariants, that fireball expansion 
should in fact  greatly enhance the sonic boom (like     
  tsunami  going onshore).

\section{QGP with magnetic quasiparticles}
  We have already mentioned that electric-magnetic duality and
Dirac quantization condition may provide a clue to why
QGP is so strongly coupled. This part of the 
 story is relatively new, in spite of many studies of magnetic
excitations on the lattice, especially in Japan.
We start with the overall picture we proposed, returning
to brief discussion of its motivation later.

{\bf The picture}
proposed is different from the traditional approach, which
puts $confinement$ phenomenon at the center of the discussion,
dividing the phase diagram into (i)
confined/hadronic phase and (ii) deconfined or QGP
 phase.
We however focus on the competition of electric and magnetic
quasiparticles, 
and divide the phase diagram  into (i) 
``magnetically dominated'' region at lower $T,\mu$ and (ii)
``electrically dominated'' one at large $T,\mu$,
separated by  ``E-M equilibrium'' line at which
 the {\em couplings}
of both interactions 
are equal\footnote{We use field theory  notations, in which
$g_e,g_m$ are electric and magnetic couplings, e/m
duality transformation is $\tau->-1/\tau$ where $\tau=\theta/2\pi+i
4\pi/g_e^2$. $g_e=g$ and $\hbar=c=1$ elsewhere.}
\be g_e^2/{4\pi\hbar c}= g_m^2/{4\pi\hbar c}= 1 \ee
The last equality follows from the
 celebrated Dirac quantization condition \cite{Dirac}
\begin{equation} \label{Dirac_quantization}
\frac{g_eg_m}{4\pi\hbar c} = \frac{n}{2}
\end{equation}
with $n$ being an integer, put to 2 because of adjoint color charge
of relevant monopoles.

\begin{figure}[t]
   \includegraphics[width=7.5cm]{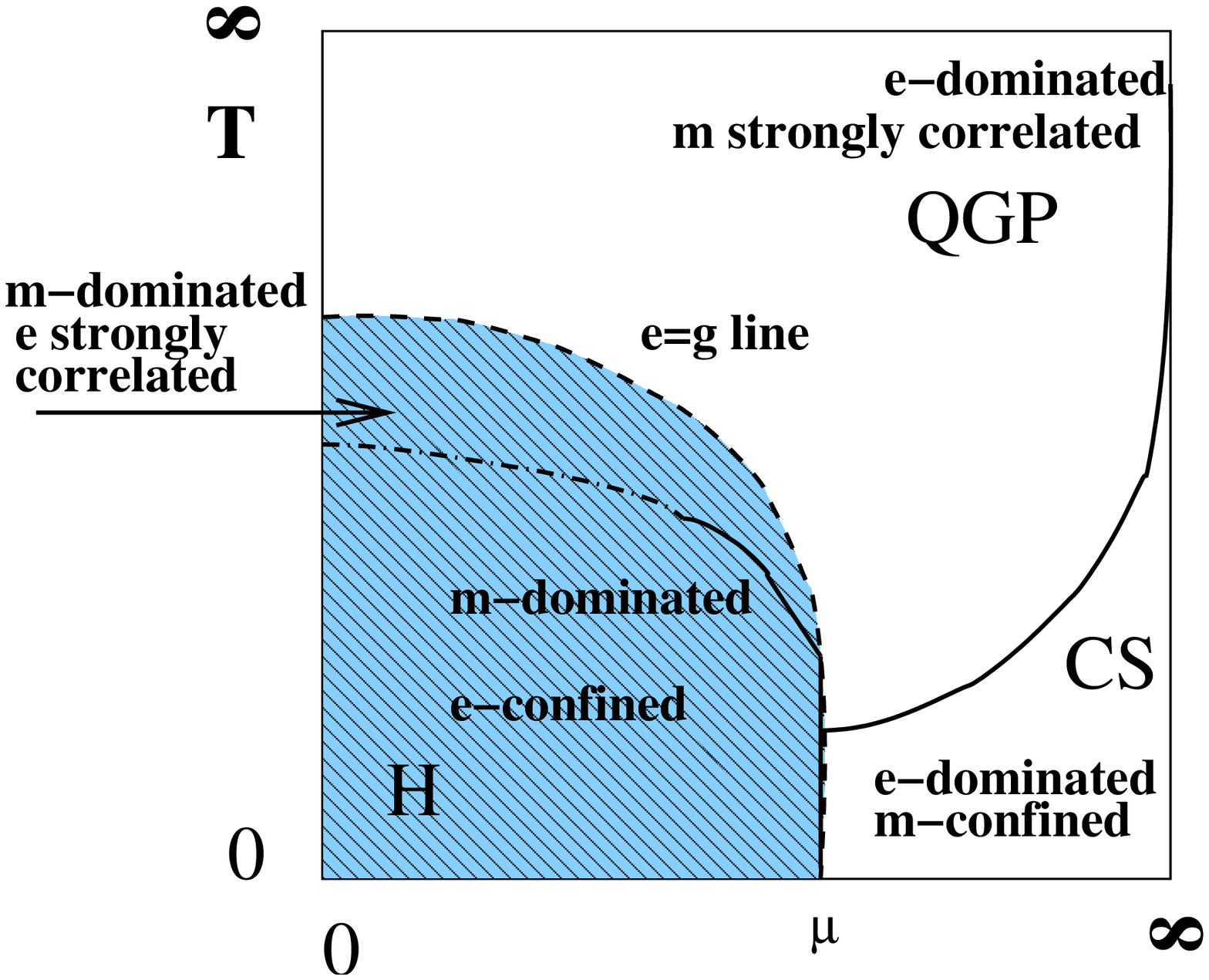}
 \includegraphics[width=7cm]{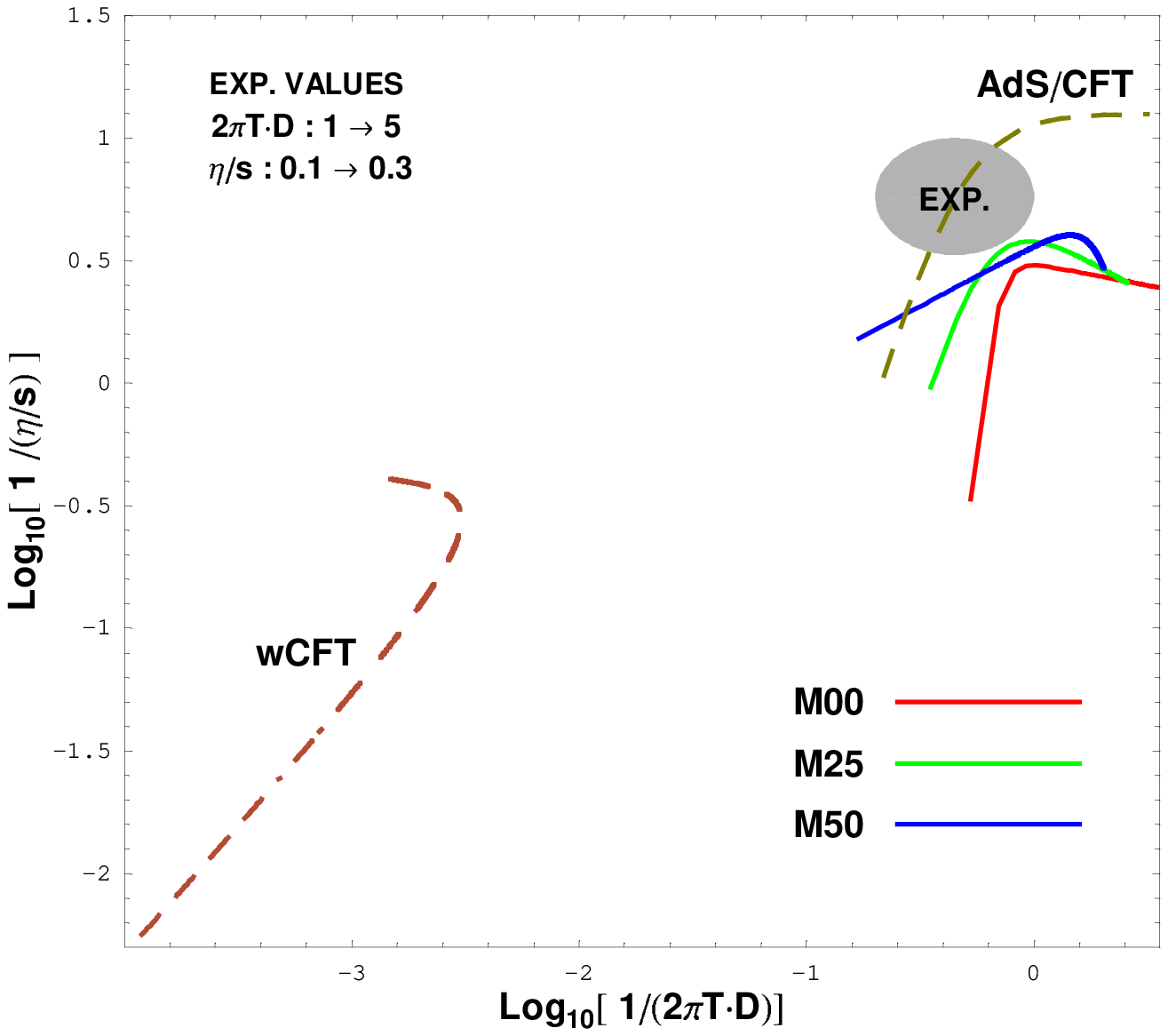}
  \vspace{0.1in}
\caption{(color online)(a) A schematic phase diagram on
 a (``compactified'') plane of
 temperature and baryonic chemical potential $T-\mu$.
 The (blue) shaded
region shows ``magnetically dominated'' region $g<e$, which
includes the e-confined hadronic phase as well as ``postconfined''
part of the QGP domain. Light region includes ``electrically
dominated'' part of QGP and also color superconductivity (CS)
region, which has e-charged diquark condensates and therefore
obviously m-confined. The dashed line called ``e=g \,line'' is the
line of electric-magnetic equilibrium.
 The solid lines indicate true phase transitions,
while the dash-dotted line is a deconfinement cross-over line. 
(b)
 Plots of $Log[1/(\eta /s)]$ v.s. $Log[1/(2\pi T D)]$ including
results from our MD simulations, the Ads/CFT calculations, the weakly coupled
CFT calculations, as compared with experimental values. M00,M25,M50
mean 0,25 and 50\% of monopoles in plasma. 
\label{fig_em_phasediag}}
\end{figure}

The ``magnetic-dominated'' low-$T$ (and low-$\mu$) region (i) can
in turn be subdivided into the $confining$ part (i-a) in which
electric field is confined into quantized flux tubes  by
magnetic condensate 
 \cite{t'Hooft-Mandelstamm}, and a new {\em ``postconfinement''}
 region (i-b) at $T_c<T<T_{E=M}$ in which electric sector is
still strongly coupled and sub-dominant
 We believe this picture better corresponds to
a situation in which string-related physics is by no means
terminated at $T=T_c$: rather it is at its maximum there. Then if
leaving this ``magnetic-dominated'' region and passing through the
equilibrium region by increase of $T$ and/or $\mu$, we enter
either the high-$T$ "electric-dominated" QGP or a (color)electric
superconductor at high-$\mu$ replacing magnetic superconductor.
(Electric diquark condensate obviously confine monopoles.)
A phase diagram explaining this pictorially is shown in
Fig.\ref{fig_em_phasediag}(a).

Besides equal couplings, the equilibrium region is also
presumably characterized by comparable densities as well as masses
of both electric and magnetic quasiparticles. In QCD the issue
is complicated by the fact that E-M duality is far from perfect,
with different spins of electric (gluons and quarks) and magnetic
quasiparticles. However other theories -- especially  $\cal N$=4
supersymmetric YM -- have perfect self-duality of electric and
magnetic description: it is also conformal and has no confinement
to complicate the picture, while E and M-dominated parameter regions
do exist\footnote{Not in the phase diagram, as in this theory
  couplings are independent of $T,\mu$.}.
see some discussion of e/m duality
in this theory in \cite{withAP}.

 Now brief motivations of this picture. One is well known
 t'Hooft-Mandelstamm scenario, in which the confined phase
 is a ``dual superconductor''. If so, there should be
be uncondensed magnetic objects above $T_c$ as well. And indeed,
 lattice studies show
that electrically charged particles -- quarks and gluons --
are getting heavier as we decrease $T$ toward $T_c$,
 while monopoles gets lighter and more numerous.\\
  The magnetic screening mass,
although absent  perturbatively, 
%
is nonzero, and even exceeds the electric one close to $T_c$  
(as shown e.g. by Nakamura et al \cite{Nakamura:2003pu}).
These screening masses as well as estimates of the densities of
electric and magnetic objects, 
leads to the location of {\em
E-M equilibrium} \cite{Liao:2006ry} at
 \be T_{E=M}\approx (1.2-1.5)T_c=250-300\, MeV\ee.


Another lattice-based puzzle 
is related with  static $\bar Q Q$ potentials close to $T_c$.
At deconfinement $T=T_c$ 
a static quark pair
has  vanishing string tension in the free energy
$exp(-F(T,r))=<W>$. However if one calculates the $energy$
or $entropy$ separately (by $F=E-TS$, $S=-\partial F/\partial T$)
one finds \cite{potentials} that
 the tension more than twice that in the vacuum,
  till rather large distances. The total energy 
added to a pair is surprisingly large,
reaching  $E(T=T_c,r\rightarrow\infty)=3-4\, GeV$, while
the entropy  $S(T=T_c,r\rightarrow\infty)\sim 20$.

Where all this energy and entropy may come from in the
 plasma phase? Most likely it is 
 due to QCD string, surviving above $T_c$ in some form. 
Liao and myself \cite{Liao:2005hj} 
suggested ``electric'' approach toward solving this puzzle,
by ``polymerization'' of gluonic quasiparticles in sQGP.
``Magnetic'' effect
is that monopoles further compress the  electric flux tube,
 as their dual -- electrons in solar plasma -- do for magnetic
flux tubes.

%

Are there bound states of electric and magnetic quasiparticles?
Yes, there are a lot of them. A surprise is that
even  finite-$T$ instantons
can be viewed as ``magnetic baryons''
being made of $N_c$ self-dual dyons~\cite{Kraan:1998kp},  attracted to
each other  pair-vise, both electrically and magnetically.
  Not only such baryons-made-of-dyons have the same moduli space
as instantons,  the solutions can be obtained
vis very interesting AdS/CFT brane
construction \cite{Lee:1997vp}. Many more exotic bound states
of those are surely waiting  to
be discovered.

 Transport properties of strongly coupled plasmas
is a non-trivial issue. Especially let us ask
 what is the role of magnetic quasiparticles, as they
may not be strongly coupled?

 These issues were
recently addressed using molecular dynamics (MD) methods of
 classical strongly coupled plasmas.
 In e/m context
the term ``strongly coupled''  is expressed via
 parameter $\Gamma= (Ze)^2/(a_{WS}T)$
characterizing the strength of the interparticle interaction.
 $Ze,a_{WS},T$ are respectively the  charge, the Wigner-Seitz radius
$a_{WT}=(3/4\pi n)^{1/3}$  and the temperature. 
Extensive studies 
using both MD and analytical methods,
have revealed the following regimes:
{\bf i.} a gas regime for $\Gamma<1$; {\bf ii.} a
liquid regime for $\Gamma\approx  10$;  {\bf iii.} a glass regime
for $\Gamma\approx 100$; {\bf iv.} a solid regime for $\Gamma > 300$. 

 Gelman, Zahed and myself~\cite{GSZ}
 proposed a  model for the description of strongly interacting
quarks and gluon quasiparticles  as a classical
and nonrelativistic $Non-Abelian$ Coulomb gas. The sign and strength
of the inter-particle interactions are fixed by the scalar product
of their classical {\it color vectors} subject to Wong's equations.


The model was studied using Molecular Dynamics (MD),
which means solving numerically EoM for $n\sim 10^2-10^3$ particles.
As the Coulomb coupling is
increased, we found at parameters corresponding to sQGP liquid-like, with a diffusion constant 
$D\approx 0.1/T$ and a bulk viscosity to entropy density ratio 
$\eta/s\approx 1/3$.
Unfortunately there is no place here to show how
 diffusion and viscosity depends on coupling: we will discuss
their interrelation in the Summary.



{\bf Plasma with magnetic charges} was studied in \cite{Liao:2006ry}
by molecular dynamics. Unlike earlier works, it does not use
periodic box but (self-contained) drops of plasma. Although
in this case the system is not homogeneous, it allows to consider
much larger systems. Simulations included transport properties such as
diffusion coefficients and viscosity. A number of collective modes have
been discovered, and their oscillation frequencies and damping
parameters calculated. The results we will show in the next section,
together with those from AdS/CFT correspondence.

\section{AdS/CFT correspondence and CFT plasma properties}

Let me omit well known results on thermodynamics,  heavy-quark
potentials
and viscosity\footnote{See Son's talk in the same proceedings.}.
%
Let me just remind that
the
 Debye radius at  strong coupling is unusual: unlike in pQCD it has no
 coupling constant.
Although potential depends on distance $r$ still
as in the Coulomb law, $1/r$ (at $T=0$ it is due to conformity),
 it is has a notorious square root of the coupling. 
Semenoff and Zarembo \cite{Semenoff:2002kk} noticed that summing ladder diagrams
one can explain $\sqrt{g^2 N_c}$, although
not a numerical constant. Zahed and myself~\cite{SZ_CFT}
pointed out that  both static charges are color correlated 
during a parametrically small time  $\delta t\sim r/{(g^2
  N_c)^{1/4}}$: this explains~\cite{Klebanov:2006jj} why a field of the dipole   
is $1/r^7$ at large distance\cite{Klebanov:2006jj}, not  $1/r^6$. 
Debye screening range can also be explained by resummation of thermal
polarizations~\cite{SZ_CFT}.
In another paper Zahed and myself~\cite{SZ_spin} had also discussed
the velocity-dependent
forces , as well as spin-spin and spin-orbit ones, at strong coupling.
Using
ladder resummation for non-parallel
Wilson lines with spin they  concluded that all of them
join into one common square root
\be V(T,r,g)\sim \sqrt{(g^2 N_c)[1-\vec v_1 *\vec v_2+(spin-spin)+(spin-orbit)]}/ r   \ee 
Here $\vec v_1,\vec v_2$ are velocities of the quarks: 
and the corresponding term is a strong coupling version of Ampere's
interaction between two currents\footnote{Note that in a 
quarkonium  their scalar product is negative,
increasing attraction.}. No results on that are known from a gravity
side.

{\bf Bound states} of fundamental particles should be present
in any strongly coupled theory\footnote{Some people publicly expressed
an opinion that they like AdS/CFT results but not multiple bound states
advocated by Zahed and myself: this is simply logically impossible,
large coupling brings both.
}.
 Zahed and myself~\cite{SZ_CFT} looked for
 heavy quarks   bound states, using a  Coulombic potential with 
Maldacena's $\sqrt{g^2 N_c}$ and
 Klein-Gordon/Dirac eqns. There is no problem with states
 at large orbital momentum $J>>\sqrt{g^2N_c}$, but otherwise one has 
the famous  ``falling on a center'' solutions\footnote{Note that all
 relativistic corrections mentioned above  cannot prevent it
from happening.}: we argued that
a significant density of bound states develops, at all energies, from zero to
$2M_{HQ}$.  

And yet, a study of the gravity side~\cite{Kruczenski:2003be} 
found that there is no falling. In more detail, the
Coulombic states at large $J$ are supplemented by two more
families: Regge ones with the mass $\sim M_{HQ}/(g^2N_c)^{1/4}$ and
 the lowest $s$-wave states
(one may call $\eta_c,J/\psi$)  with even smaller
masses  $\sim M_{HQ}/\sqrt{g^2  N_c}$.
 The issue of ``falling'' was further discussed
 by Klebanov, Maldacena and Thorn~\cite{Klebanov:2006jj} for
 a pair of static quarks: they calculated the
 spectral density of states
  via a semiclassical quantization of  string vibrations.
They argued  that their corresponding density of states
should appear at exactly the same
critical
coupling as the famous ``falling'' in the Klein-Gordon eqn..

AdS/CFT also has  multi-body states  similar
to   ``polymeric
chains'' $\bar q.g.g... q$ discussed for sQGP in\cite{Liao:2005hj}. 
 Hong, Yoon and Strassler~\cite{Hong:2004gz} have studied
such states when
the endpoints are static quarks, and the intermediate
gluons  are conveniently replaced by adjoint scalars,
so that one can use their ``flavor'' to see how long the chain is.

{\bf Heavy quark transport } in
 the CFT plasma 
was a subject of recent breakthroughs.
Heavy quark diffusion constant
has been calculated by Casalderrey-Solana and
Teaney~\cite{Casalderrey-Solana:2006rq}: 
\be D_{HQ}={2 \over \pi T \sqrt{g^2 N_c}}  \ee
which leads to stopping length much  smaller than 
an expression for the momentum diffusion
$D_p=\eta/(\epsilon+p)\sim 1/4\pi T$.
This work is methodically quite different from others: one has
to use full Kruskal
coordinates, including the inside
of the black hole connecting
 $two$ Universes (with opposite time directions)
simultaneously, see Fig.3a\footnote{So to say, a fish and a fisherman 
(corresponding to an amplitude and
a conjugated amplitude) live separately in these Universes, yet
they are connected by a string which conducts a  wave -- an
 information about a ``bite''.} 
Further important result \cite{Casalderrey-Solana:2007qw}
is calculation of the mean transverse energy squared per unit length
(1/2 of the popular parameter $\hat q$ for a gluon)
for a quark moving with gamma factor $\gamma=1/\sqrt{1-v^2}$ 
\be k_t=\sqrt{\gamma \lambda}\pi T^3\ee
  
Jet quenching
 studies by many authors\cite{jq}
 have resulted in
 the following expression for the drag force
\be {dP\over dt}= -{\pi T^2\sqrt{g^2 N_c} v \over  2\sqrt{1-v^2}} \ee
Quite remarkably, the Einstein relation which 
relates  the heavy quark diffusion constant (given above)
to the drag force
is actually fulfilled, in spite of quite different gravity settings.
This result is valid 
 in a stationary setting, in which
a quark is dragged with constant velocity $v$ by ``an invisible
hand'',
 see Fig.3b.  Friess et al
\cite{Friess:2006aw} then 
 solved linearized
 Einstein equations 
and found corrections to the metric
$h_{\mu\nu}$ induce by a falling string,
obtaining the stress tensor of floating  matter
on the brane. Quite remarkably, when they analyzed harmonics
of this stress at small momenta they have found the ``conical flow'',
disappearing at ``subsonic'' $v<1/\sqrt{3}$ and peaked at
the Mach cone. Recent studies of near-zone  flow
\cite{Gubser:2007nd} found that picture changes at velocity
a bit above the speed of sound, and a different
angle of the maximum. This indeed should
 happen because finite amplitude
shocks do propagate with a speed $larger$ than that of sound.

\begin{figure}
\includegraphics[width=6.1cm]{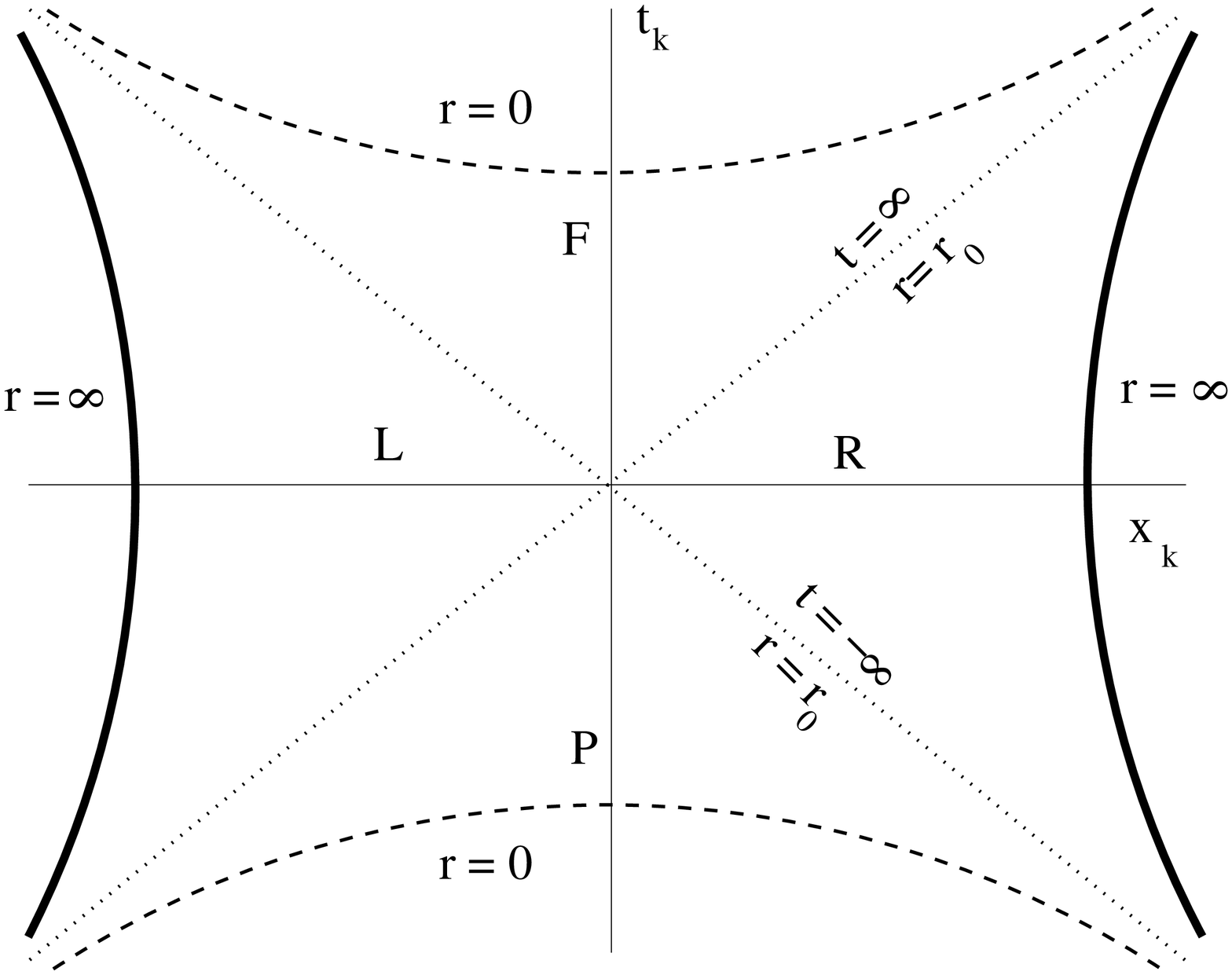}
\includegraphics[width=7cm]{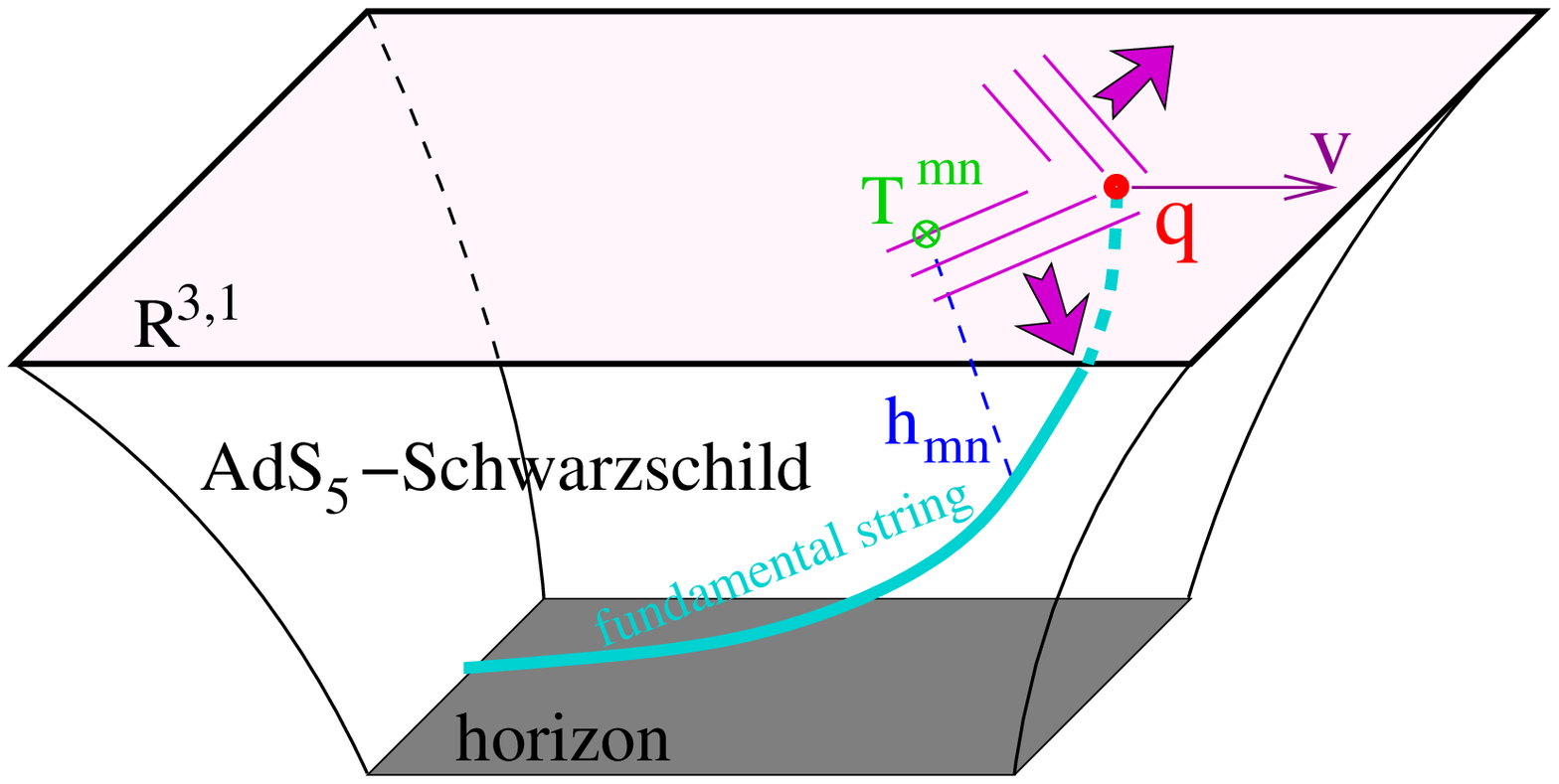}
 \caption{\label{fig_wake}
(a) (from \protect\cite{Casalderrey-Solana:2006rq}): In Kruskal coordinates one can study two Universes
at the same time, shown right and left, and the evaluated
Wilson line contains static quarks on their boundaries.
(b) (from \protect\cite{Friess:2006aw})
The dragged quark trails a string into the five-dimensional AdS
 bulk, representing color fields sourced by the quark's fundamental
 charge and interacting with the thermal medium. The back gravity
 reaction describes how matter flows on the brane.
 }
 \end{figure}


  {\bf ``Gravity duals''} to complete hydrodynamical explosion is  
perhaps  the ultimate  AdS/CFT application:
but unfortunately solving  non-linearized
Einstein equations with evolving black holes
is a very challenging task, both conceptually and technically.


The need for
{\em dynamically  generated}  black hole (BH),
corresponding to thermal fireball,
 was suggested  by Nastase~\cite{Nastase}.
 Sin, Zahed and myself~\cite{Shuryak:2005ia}
 pointed out that exploding/cooling
fireball on the brane is dual to
 $departing$  black hole, 
falling toward the AdS center. Although specific solution 
 discussed in that paper 
has a brane departing from a static black hole, corresponding to
spherically symmetric solution with a time-dependent
$T$, we also 
suggested three d-dimensional toy problems, 
 corresponding for d=1 to
a collision of two  walls and subsequent Bjorken
rapidity-independent expansion, with 
 2d and 3d corresponding to cylindrical and spherical 
collapses.
Janik and Peschanski~\cite{JP} soon
 found asymptotic (late-time) solution corresponding to
 d=1 rapidity-independent Bjorken expansion. It indeed has   
 a departing $horizon$ at the 5-th distance $z_h\sim \tau^{1/3}$
  Further discussion of the first subleading terms $O(\tau^{-2/3})$ 
has been made by Sin and Nakamura  \cite{SinNak}  who 
 identified them with the viscosity effects, although without
any preference for its particular value.
 Janik \cite{janik2} 
derived next sub-sub-leading term  $O(\tau^{-4/3})$
and provided some argument why viscosity should be what
was found in equilibrium. 
Lin and myself\cite{Lin_Shu} have discussed a process of
BH formation, by studying motion of various objects in
the 5-th direction. 
They concluded that all  strings, closed or open, quickly 
form 
a thin shell (or 3d-membrane) of matter, which is gravitationally
 collapsing toward the AdS center, and related its EoM to
Israel junction equations
for the motion of collapsing shell. 

  Princeton group  \cite{Friess:2006kw} 
 pointed out that global static black-hole-AdS metric can
be seen from viewpoint of a moving observer, which 
is nevertheless living in asymptotically flat Minkowskian world.
This provides nice analytic solution with incoming and then 
departing 
BH, corresponding to d=3
 spherical
explosion at the boundary. 
 This explosion is however exactly reversible,  conserving
the entropy.
Elliptic deformation of this solution leads to
hydrodynamics with dissipation. First the value
of the damping for corresponding modes have $not$
matched the expected CFT viscosity, but this was then corrected
 by Michalogiorgakis and Pufu, and Siopsis
 \cite{new_boundary_cond} who pointed out
that the usual Dirichlet condition 
on the boundary are to be modified.

\section{Summary of transport properties}

Finally, we would like to compare our results with those
obtained using the AdS/CFT correspondence and also with empirical data
about sQGP from RHIC experiments. Those are summarized in
Fig.\ref{fig_em_phasediag}(b), as a log-log plot of properly normalized
dimensionless (heavy quark) diffusion constant and viscosity.

The dashed curve in the left lower corner is for $\cal N$=4 SUSY
YM theory in weak coupling, where viscosity is from
\cite{Huot:2006ys} and diffusion constant
from~\cite{Chesler:2006gr}. The curve has a slope of one on this
plot, as in weak coupling both quantities are proportional to the
same mean free path. Note that weak coupling results
are quite far from empirical data from RHIC, shown by a gray oval
in the right upper corner. Viscosity estimates follow from
deviations of the elliptic flow at large $p_t$ from hydro
predictions 
while diffusion constants are
estimated from $R_{AA}$ and elliptic flow of charm
\cite{MT}.

The curve for strong-coupling AdS/CFT results (viscosity according
to \cite{PSS} with $O(\lambda^{-3/2})$ correction, diffusion
constant from \cite{Casalderrey-Solana:2006rq}), shown by upper dashed
line, is on the other hand going right through the empirical
region. At infinite coupling this curve reaches $s/\eta=4\pi$
which is conjectured to be its bound.
Our MD results\cite{Liao:2006ry} -- three solid lines on the right -- correspond to our
calculations with different ratio of electric/magnetic quasiparticles.
The overall behavior of these so different approaches,
as well as proximity to the empirical range, is very encouraging.

\section*{Acknowledgements}
I thank the organizers of the meeting and
 Yukawa Institute for Theoretical Physics at Kyoto University for
 hospitality.
My research
 is partially supported by the US-DOE grants DE-FG02-88ER40388
and DE-FG03-97ER4014.


\begin{thebibliography}{99}


\bibitem{Shu_liquid}
  E.V.Shuryak,
  Prog.\ Part.\ Nucl.\ Phys.\  {\bf 53}, 273 (2004)
  [ hep-ph/0312227].

\bibitem{SZ_rethinking}
E.V.Shuryak and I. Zahed, {\tt hep-ph/0307267},
Phys.\ Rev.\ C {\bf 70}, 021901 (2004)
Phys.\ Rev.\ D {\bf 70}, 054507 (2004), hep-ph/0403127.
\bibitem{SZ_CFT}
E.V.Shuryak and I. Zahed,
Phys.\ Rev.\  {\bf D69} (2004) 014011.
[ hep-th/0308073].

\bibitem{MT}
  G.~D.~Moore and D.~Teaney,
   hep-ph/0412346.



\bibitem{charmonium}
S.~Datta, F.~Karsch, P.~Petreczky and I.~Wetzorke,
{\tt hep-lat/0208012}. 
M. Asakawa and T. Hatsuda,
Nucl. Phys. {\bf A715} (2003) 863c; 
hep-lat/0308034;
\bibitem{Rapp_vanHees}
  H.~van Hees, V.~Greco and R.~Rapp,
  hep-ph/0601166.

\bibitem{Gelman:2004fj}
  B.~A.~Gelman, E.~V.~Shuryak and I.~Zahed,
 Phys.Rev. A 72, 043601 (2005) 
\bibitem{GSZ} B.~A.~Gelman, E.~V.~Shuryak and I.~Zahed,
Phys.\ Rev.\  C {\bf 74}, 044909 (2006) ,
  nucl-th/0601029,
nucl-th/0605046.
\bibitem{Liao:2006ry}
  J.~Liao and E.~Shuryak,
  hep-ph/0611131.
\bibitem{CST}
  J.~Casalderrey-Solana, E.~V.~Shuryak and D.~Teaney,
   hep-ph/0411315.
  hep-ph/0602183.
\bibitem{Antinori:2005tu}
  F.~Antinori and E.~V.~Shuryak,
   nucl-th/0507046.
\bibitem{CS_variable}J.~Casalderrey-Solana and E.~V.~Shuryak,
  hep-ph/0511263.
\bibitem{Dirac}
P. A. M. Dirac, Proc. R. Soc. London {\bf A133}, 60 (1931).
\bibitem{t'Hooft-Mandelstamm}
S.~Mandelstam,
  Phys.\ Rept.\  {\bf 23}, 245 (1976).\\
G.~'t Hooft,
   ``Topology Of The Gauge Condition And New Confinement Phases In Nonabelian
  Nucl.\ Phys.\ B {\bf 190}, 455 (1981).


\bibitem{Kraan:1998kp}
  T.~C.~Kraan and P.~van Baal,
  Phys.\ Lett.\ B {\bf 428}, 268 (1998)
  [hep-th/9802049].



\bibitem{withAP} A.Parnachev and E.Shuryak, Electric-magnetic
duality and transport properties, in QCD and AdS/CFT. In progress.  

 \bibitem{Nakamura:2003pu}  A.~Nakamura, T.~Saito and S.~Sakai,
  Phys.\ Rev.\ D {\bf 69}, 014506 (2004)
  [hep-lat/0311024].
 \bibitem{masses}
 P. Petreczky, F. Karsch, E. Laermann, S. Stickan, I. Wetzorke,
 Nucl. Phys. Proc. Suppl. {\bf 106} (2002) 513.
 \bibitem{LS} J.~Liao and E.~V.~Shuryak,
Nucl.Phys.A, in press, hep-ph/0508035.
  Phys.\ Rev.\ D {\bf 73}, 014509 (2006)
  [hep-ph/0510110].


\bibitem{Shuryak:2006ap}
  E.~V.~Shuryak,
  nucl-th/0606046.

\bibitem{Semenoff:2002kk}
  G.~W.~Semenoff and K.~Zarembo,
  Nucl.\ Phys.\ Proc.\ Suppl.\  {\bf 108}, 106 (2002)
  [hep-th/0202156].
\bibitem{Klebanov:2006jj}
  I.~R.~Klebanov, J.~M.~Maldacena and C.~B.~Thorn,
  JHEP {\bf 0604}, 024 (2006)
  [hep-th/0602255].
  C.~G.~.~Callan and A.~Guijosa,
  Nucl.\ Phys.\ B {\bf 565}, 157 (2000)
\bibitem{SZ_spin}
  E.~V.~Shuryak and I.~Zahed,
  Phys.\ Lett.\ B {\bf 608}, 258 (2005)
  [hep-th/0310031].
\bibitem{Kruczenski:2003be}
  M.~Kruczenski, D.~Mateos, R.~C.~Myers and D.~J.~Winters,
  JHEP {\bf 0307}, 049 (2003)
  [hep-th/0304032].

\bibitem{Hong:2004gz}
  S.~Hong, S.~Yoon and M.~J.~Strassler,
  JHEP {\bf 0603}, 012 (2006)
  [hep-th/0410080].


\bibitem{PSS}
G.~Policastro, D.~T.~Son and A.~O.~Starinets,
Phys.\ Rev.\ Lett.\  {\bf 87} (2001) 081601.

\bibitem{Casalderrey-Solana:2006rq}
J.~Casalderrey-Solana and D.~Teaney,
 {{\tt hep-ph/0605199}}.


\bibitem{jq}
%
C.~P. Herzog, A.~Karch, P.~Kovtun, C.~Kozcaz, and L.~G. Yaffe, 
{{\tt hep-th/0605158}}.
%
S.~S. Gubser, 
 {{\tt hep-th/0605182}}.
A.~Buchel, 
 {{\tt
  hep-th/0605178}}.
S.-J. Sin and I.~Zahed, 
  {{\tt hep-ph/0606049}}.

\bibitem{Friess:2006aw}
J.~J. Friess, S.~S. Gubser, and G.~Michalogiorgakis, 
  {{\tt hep-th/0605292}}.
\bibitem{new_boundary_cond}
G.~Michalogiorgakis and S.~S.~Pufu,
  hep-th/0612065.
G.~Siopsis,
  hep-th/0702079.


\bibitem{Nastase}H.~Nastase,
  hep-th/0501068.
\bibitem{Shuryak:2005ia}
  E.~Shuryak, S.~J.~Sin and I.~Zahed,
  hep-th/0511199.

\bibitem{JP}
  R.~A.~Janik and R.~Peschanski,
 Phys.\ Rev.\ D {\bf 73}, 045013 (2006)
  hep-th/0512162,
  hep-th/0606149.



%
\bibitem{Liao:2005hj}
  J.~Liao and E.~V.~Shuryak,
  Nucl.\ Phys.\  A {\bf 775}, 224 (2006)
  [hep-ph/0508035].

\bibitem{potentials}
O.~Kaczmarek, S.~Ejiri, F.~Karsch, E.~Laermann and F.~Zantow,
{\tt hep-lat/0312015}.


\bibitem{Lee:1997vp}
  K.~M.~Lee and P.~Yi,
  Phys.\ Rev.\ D {\bf 56}, 3711 (1997)
  [hep-th/9702107].

\bibitem{Lin_Shu}
  S.~Lin and E.~Shuryak,
  hep-ph/0610168.

\bibitem{Friess:2006kw}
  J.~J.~Friess, S.~S.~Gubser, G.~Michalogiorgakis and S.~S.~Pufu,
  hep-th/0611005.

\bibitem{JP}
  R.~A.~Janik and R.~Peschanski,
  Nucl. Phys.  {\bf B586} (2000) 163.
\bibitem{SinNak}   S.~Nakamura and S.~J.~Sin,
  hep-th/0607123.
 \bibitem{Israel}W.Israel, Nuovo Cimento XLIV B,1 (1966)
\bibitem{janik2}
  R.~A.~Janik,
  Phys.\ Rev.\ Lett.\  {\bf 98}, 022302 (2007)
  [hep-th/0610144].

\bibitem{Chesler:2006gr}
  P.~M.~Chesler and A.~Vuorinen,
  JHEP {\bf 0611}, 037 (2006)
  [hep-ph/0607148].

\bibitem{Huot:2006ys}
  S.~C.~Huot, S.~Jeon and G.~D.~Moore,
  hep-ph/0608062.


\bibitem{Casalderrey-Solana:2007qw}
  J.~Casalderrey-Solana and D.~Teaney,
  hep-th/0701123.

\bibitem{Gubser:2007nd}
  S.~S.~Gubser and S.~S.~Pufu,
  hep-th/0703090.
A.~Yarom,
  hep-th/0703095.
\end{thebibliography}
\end{document}